\documentclass{raa}
\usepackage{graphicx,times}
\usepackage{amssymb,amsmath}
\usepackage[colorlinks=true,linkcolor=green,citecolor = blue]{hyperref}
\begin{document}

\title{Physical and geometrical parameters of CVBS XIV: The two nearby systems HIP 19206 and HIP 84425}
%  \subtitle{Physical and geometrical parameters of CVBS XIV: HIP 19206 and HIP 84425}
 \volnopage{Vol.0 (20xx) No.0, 000--000}      %%preserved for Editor. DOn't remove!
 \setcounter{page}{1}          %%starting page, preserved for Editor. DOn't remove!

\author{Al-Wardat  M. A.\inst{1,2}, Abu-Alrob E. \inst{3}, Hussein A. M. \inst{3}, Mardini M. \inst{4}, Taani A. A. \inst{5}, Widyan H. S. \inst{3},  Yousef Z. T.\inst{6},    Al-Naimiy H. M. \inst{1,2},  Yusuf N. A. \inst{7}}
   \institute{Department of Applied Physics and Astronomy, University of Sharjah, Sharjah, UAE\\
e-mail:malwardat@sharjah.ac.ae\\
\and
 Sharjah Academy for Astronomy, Space Science and Technology, Sharjah, UAE\\
\and
Department of Physics, Al al-Bayt University, Mafraq, 25113 Jordan\\
\and
 Key Lab of Optical Astronomy, National Astronomical Observatories, Chinese Academy of Sciences, Beijing 100102, China\\
 \and
 Physics Department, Faculty of Science, Al-Balqa Applied University, 19117 Salt, Jordan\\
 \and
 Department of Applied Physics, Faculty of Science and Technology, Universiti Kebangsaan Malaysia, 43600 UKM Bangi, Selangor, Malaysia \\%e-mail:zahraa-k@hotmail.com\\
 \and
Department of Physics, Yarmouk University, Irbid, 21163, Jordan\\
%e-mail:nihadyusuf@yu.edu.jo
\vs\no
   {\small Received~~20xx month day; accepted~~20xx~~month day}}

\abstract{
The data release DR2 of Gaia mission was of great help in  precise determination of fundamental parameters of Close Visual Binary and Multiple Systems (CVBMSs), especially masses of their components, which are crucial parameters in understating formation and and evolution of stars and galaxies.
This article presents the complete set of fundamental  parameters of two nearby  (CVBSs), these are HIP 19206 and HIP 84425.
We used a combination of two methods; the first one is Tokovinin's dynamical method to solve the orbit of the system and to estimate orbital elements and the dynamical mass sum, and the second one is Al-Wardat's method for analyzing CVBMSs to estimate the physical parameters of the individual components. The latest method   employs grids of Kurucz line-blanketed plane parallel model atmospheres to build  synthetic Spectral Energy Distributions (SED) of the individual components.
Trigonometric parallax measurements given by Gaia DR2 and Hipparcos catalogues are used to analyse the two systems. The difference in these measurements  yielded slight discrepancies in the fundamental parameters of the individual components especially masses. So, a new dynamical parallax is suggested in this work based on the most convenient mass sum as given by each of the two methods. The new dynamical parallax for the system Hip 19205  as $22.97\pm 0.95$ mas coincides well with the trigonometric one given recently (in December 2020) by Gaia DR3  as $22.3689\pm 0.4056$ mas.
 The positions of the components of the two systems  on the evolutionary tracks and isochrones are plotted, which suggest  that  all components are solar-type main sequence stars. Their most probable formation and evolution scenarios are also discussed.
}

   \authorrunning{M. A. Al-Wardat et al. }            %author_head in even pages
   \titlerunning{Physical and geometrical parameters of CVBS XIV: HIP 19206 and HIP 84425}  % title_head in odd pages
   \maketitle
\section{Introduction}
\label{sect:intro}
The study of the physical and geometrical  parameters, especially masses, of close visual binary systems (CVBSs) plays an important role in revealing the secrets of formation and evolution of such systems, and hence the formation and evolution of the Galaxy.
It also represents a direct tool for the investigation of laws and rules of astrophysics.

The preciseness of such studies is highly enhanced by the latest measurements of the distances to such systems, which were highly achieved by Gaia astrometric mission \cite{gaia2018vizier, al-wardat_hussein_al-naimiy_barstow_2021}.

The two systems under study fulfill the following specific requirements;  The two components of the system are visually close enough that they can not be observed and studied individually, and the system has available accurate  measurements of its colours, colour indices and  magnitude difference between its  components.

The study uses two complementary methods; Tokovinin's dynamical method for the orbital solutions, and Al-Wardat's complex method for analyzing CVBMSs.

Al-Wardat's method is an indirect computational spectrophotmetrical method that uses the available observations (magnitudes and magnitude differences) of a binary or a multiple system to build a synthetic Spectral Energy Distribution (SED) for each individual component. It employs grids of Kurucz's line-blanketed plane parallel atmospheres to build these synthetic SEDs ~\cite{1994KurCD..19.....K}. Then it combines these SEDs depending on some geometrical information like the system's parallax and the radii of the components to get the entire SED of the system, from which we can calculate the synthetic magnitudes and colour indices.  This process is  subject to iteration until the best fit between the synthetic and observational magnitudes and colour indices are achieved. Whereupon, input parameters,  both those of the model atmospheres and the geometrical calculations, represent the fundamental parameters of the system adequately enough within the error values of the measured ones.

This methods was successfully applied to several CVBMSs.  Some of them were solar type stars, while the others were found to be  subgiant stars.

The criterion of the judgement of Al-Wardat's method is the best fit between the observational entire  colours and colour indices with the synthetic ones.
In the case of the existence of the entire observational SED, it is usually used as an additional judgement tool by achieving the best fit with the synthetic one.

Of the solar type CVBSs which studied using Al-Wardat's method we mention here: ADS 11061, COU 1289, COU 1291, HIP 11352, HIP 11253, HIP 70973, HIP 72479,  Gliese 150.2, Gliese 762.1, COU 1511, FIN\,350, and HIP 109951 (see i.e. \cite{2002BSAO...53...51A, 2007AN....328...63A, 2009AN....330..385A, 2012PASA...29..523A, 2014AstBu..69...58A, al2016physical, masda2016physical, al2017physical, masda2019physical, 2019AstBu..74..464M}.
Of course, the method can deduce if the system is evolved or not!, here are some evolved (sub-giants) CVBSs which were analyzed using the method; HD\,25811,  HD\,375 and HD\,6009 (see i.,e.  \cite{2014PASA...31....5A, 2014AstBu..69..454A, 2014AstBu..69...58A} for full details and references).

This paper  presents the analysis of two nearby CVBSs; these are HIP 19206 (HD\,26040) and HIP 84425 (HD\,155826).

\section{Archived data}
The observational photometric data are taken from reliable different sources, such as the old Hipparcos Catalogue \cite{1997yCat.1239....0E}, new Hipparcos reduction \cite{2007A&A...474..653V} and Gaia data release 2  (DR2)  \cite{gaia2018vizier}.

\begin{table}[ht]
	\centering
	\caption{Observational data for HIP 19206 and HIP 84425.} \label{tab1}
	\begin{tabular}{cccc}\hline\hline\
	Hip &  19206 &	84425 & Source of data \\
	HD	& 26040 & 155826 & \\
	
		\hline
		$\alpha_{2000}$&   $04^{\rm h} 07^{\rm m} 00^{\rm s}566$&
		$17^{\rm h}15^{\rm m}35^{\rm s}902$&  SIMBAD$^1$\\
		$\delta_{2000}$ &$-10\degr00' 01.''857$ & $-38\degr35' 38.''902$ & -\\
		Sp. Typ.  &  G1/2V& G0V  &-		\\
		$A_V$       &  $0.0$ &$0.0062$ &  \cite{2018AA...616A.132L}		\\
		$V_J(Hip)$  & 6.88 & 5.95& \cite{1997yCat.1239....0E}		\\
			$(V-I)_J$&   $0.65\pm0.00$ &$0.64\pm0.02$ & -		\\				$(B-V)_J$&   $0.576\pm0.004$ & $0.580\pm0.004$ &-	\\
		    $B_T$&$7.566\pm0.008$ & $6.67\pm0.013$ &-	\\			
		    $V_T$  & $6.957\pm0.006$ & $6.015\pm0.006	$ &-		\\
		$\pi_{Hip}$ (mas)      & $24.00\pm0.92$& $32.6\pm1.7$ & -		\\
		$\pi_{Hip}$ (mas)    &  $24.49\pm0.83$ & $32.69\pm0.59$ & Van Leeuwen 2007		\\
		$\pi_{Gaia}$ (mas)      & $23.4039\pm0.4727$ & $30.4039\pm0.4728$ & \cite{gaia2018vizier}\\
		\hline\hline
	\end{tabular}
	\\
	Notes.\\
	$^1$ http://simbad.u-strasbg.fr/simbad/sim-fid.\\
\end{table}

 These data are used to calculate the preliminary input parameters and as reference for the best fit with the synthetic  photometry. Table~1 contains the catalogued  data for  HIP 19206 and HIP 84425. These data were taken  from SIMBAD database, the Hipparcos and Tycho Catalogues  and Gaia DR2.  For the orbital solutions, we used data from the Fourth Catalog of Interferometric Measurements of Binary Stars \cite{2001AJ....122.3480H}.
 \section{Methods and Analysis}

\subsection{Orbital solutions}\label{22}

\begin{figure}
	\centering
	\includegraphics[angle=0,width=12.0cm]{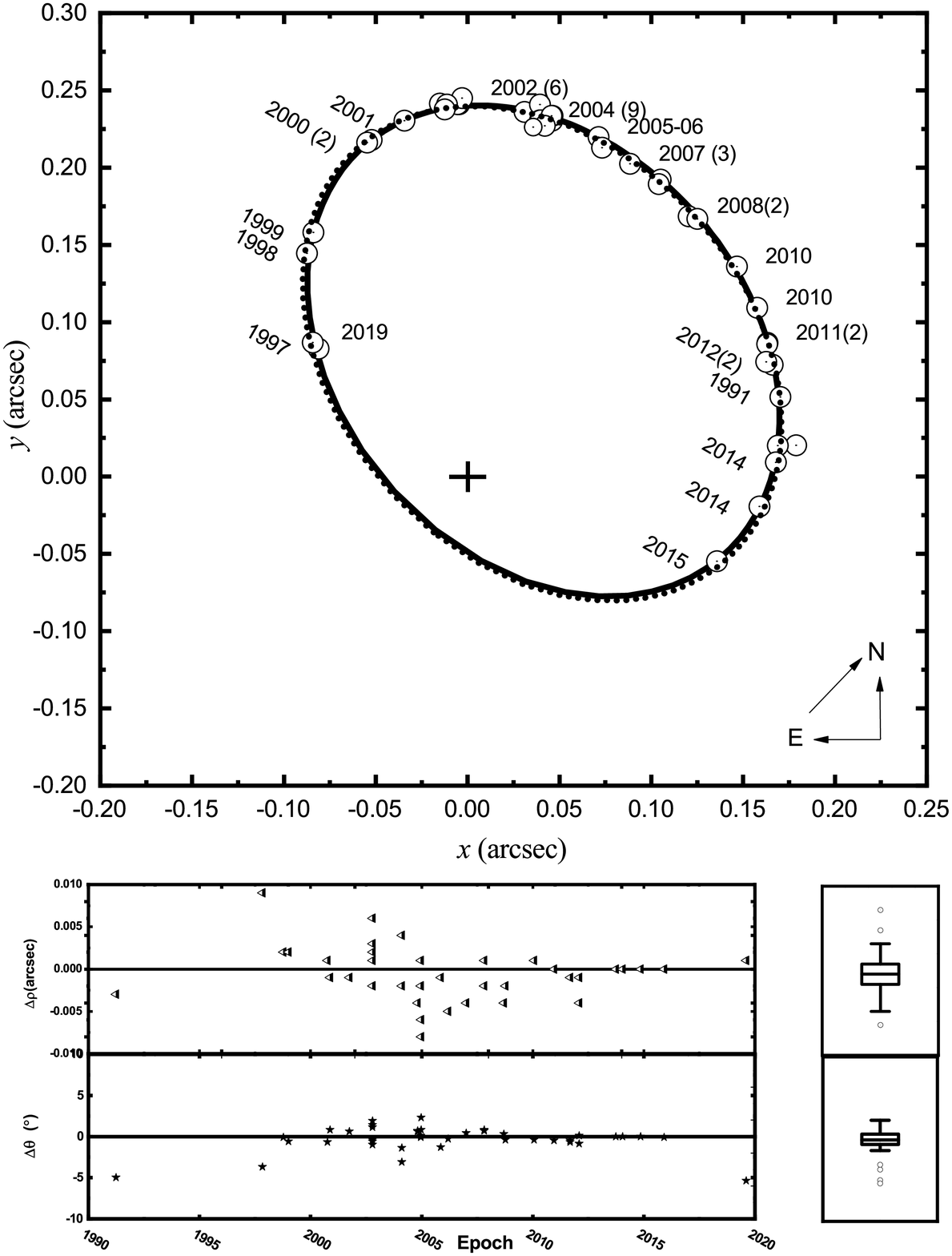}
	\caption{The modified orbit of the HIP 19206 system as determined from our analysis (solid line) and last orbit from ORB6 (dotted line). The bottom left panel shows the fit residuals, showing the difference between the observed and model values for the angular separation ($\Delta \rho $) and position angle ($\Delta \theta$) of the orbit. The bottom right panel represents the distribution of data based on a five-number summary ("minimum" first quartile (Q1); median, third quartile (Q3); and "maximum") and outliers.}
		\label{orb19206}
\end{figure}

\begin{figure}
	\centering
	\includegraphics[angle=0,width=12.0cm]{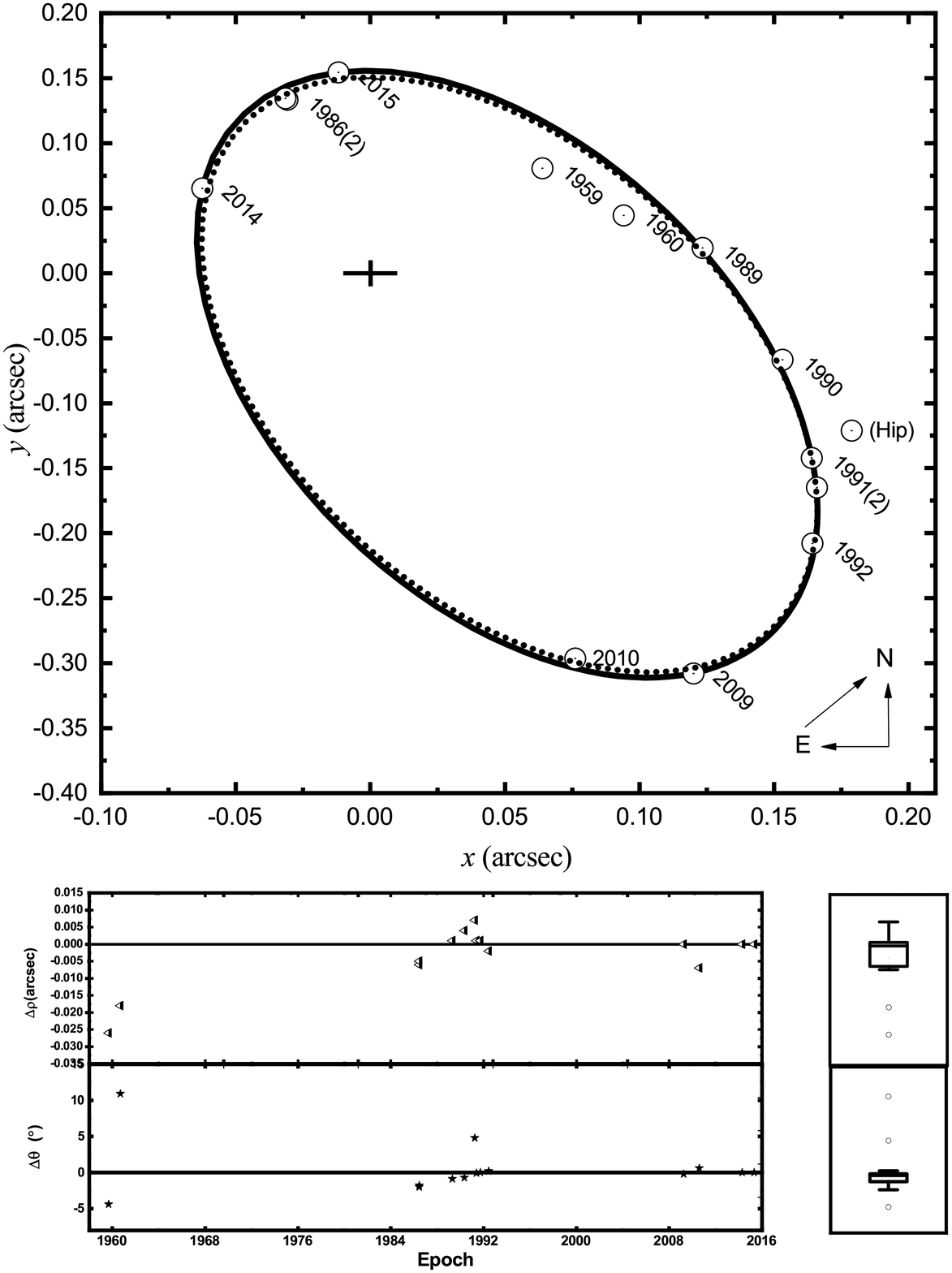}
	\caption{The modified orbit of the HIP 84425 system as determined from our analysis (solid line) and last orbit from ORB6 (dotted line). The bottom left panel shows the fit residuals, showing the difference between the observed and model values for the angular separation ($\Delta \rho $) and position angle ($\Delta \theta$) of the orbit. The bottom right panel represents the distribution of data based on a five-number summary ("minimum" first quartile (Q1); median, third quartile (Q3); and "maximum") and outliers.}
		\label{orb844425}
\end{figure}

For the orbital solution and  modification of the orbital elements of the systems,  we follow Tokovinin's dynamical method using the ORBITX code of \cite{2016AJ....151..153T}.  We used the relative position measurements; angular separations ($\rho$) in ($\arcsec$) and position angles ($\theta$) in ($\deg$) obtained by different  observational techniques and listed in the  Fourth Catalog of Interferometric Measurements of Binary Stars (INT4). The program performs a least-squares adjustment to all available relative position observations, with weights inversely proportional to the square of their standard errors.

The orbital solution involves the orbital period $P$ (in years); the eccentricity e, the semi-major axis $a$ (in arcsecond), the inclination $ \rm i$ (in degree), the argument of periastron $\rm\omega$ (in degree), the position angle of nodes $\rm \Omega$ (in degree), and the time of periastron passage $\rm T_0$ (in years).

\textbf{\,HIP 19206}; the Preliminary orbital parameters of the system were calculated by \cite{2006AA...448..703B}, then modified two times by \cite{2015AJ....150...50T} and \cite{2019IAUDS.199....1C}.

In this work, we use 41 points of positional measurements ($\rho$) and ($\theta$) covers the period from  1991 to  2019. The latest points are published by: \cite{2015AJ....150...50T} (one point), \cite{2018yCat..51550215M}(3-points), and \cite{2019IAUDS.199....1C} (one point).
\begin{table}[ht]
    \centering
    \caption{Modified orbital elements,  mass sums, and quality controls of the system HIP 19206 obtained in this work, along with previous ones of: (1) \cite{2006AA...448..703B}, (2) \cite{2015AJ....150...50T}  and (3) \cite{2019IAUDS.199....1C}.}

    \label{tab33}
    \begin{tabular}{cccccc}
    \noalign{\smallskip}
			\hline\hline
			\noalign{\smallskip}
			&	&\multicolumn{4}{c}{System HIP 19206} \\
			\cline{3-6}
			\noalign{\smallskip}
			Parameters	& Units&(1) & (2)& (3) &This work\\
			\hline
			$ P \pm \sigma_{ P}$ & [yr]  & $21.33\pm 0.44$      & $21.42\pm 0.037$ &21.06 &$21.103\pm 0.00$
			\\
			$ T_0 \pm \sigma_{ T_0}$ & [yr] & $1996.77 \pm 0.08$  & $1996.78 \pm 0.06$& 2017.94& $1996.91 \pm 0.00$
			\\
			$ e \pm \sigma_{\rm e}$  & - & $0.686 \pm 0.011$  & $0.687 \pm 0.009$ & 0.711& $ 0.7019   \pm 0.00$\\
			$ a \pm \sigma_{\rm a}$& [arcsec] & $0.223\pm 0.006 $& $0.225\pm 0.006$ & 0.229 & $0.226\pm 0.000$
			\\
			$ i \pm \sigma_{\rm i}$ & [deg]  & $ 122.2\pm 1.2 $    & $ 122.9\pm 1 $& 121.7 &$122.68\pm 0.00$ \\
			$ {\Omega}\pm \sigma_{\rm{\Omega}}$  & [deg]   &  $255.6\pm 1.1$ & $219.0\pm 1.8$& 220.4 & $219.00 \pm 0.08$\\
			$ \omega \pm \sigma_{\rm \Omega} $& [deg]& $38.2 \pm 1.7$ & $76.4 \pm 0.5$& 77.3 &$76.40 \pm 0.05$ \\
			
			$\rm M_{dyn} {\pm\sigma_{\rm M}}^{1}$ & [M$_\odot$]  & $ 1.763\pm0.258$ & $1.796\pm0.252$& $1.959$& $1.873\pm0.215$\\
			
			$\rm M_{dyn} {\pm\sigma_{\rm M}}^{2}$ & [M$_\odot$] &$ 1.659\pm0.226$&$1.690\pm0.219$& $1.843$ & $1.762\pm0.179$\\
				
			$\rm M_{dyn}{\pm\sigma_{\rm M}}^{3}$ & [M$_\odot$]  &$ 1.965 \pm0.215 $  &$ 2.001\pm0.202$& $2.183$ & $2.087\pm0.128$\\
			
			\hline
			$rms (\theta)$ & [deg] & 0.9 & 0.2335 & & 0.10
			\\
			$rms (\rho)$ & [arcsec] & 0.002 & 0.9268 & &0.0002
			\\
			\hline\hline
			\noalign{\smallskip}
		\end{tabular}\\
			Notes.\\
		$^{1}$ Using ($\pi_{Hip1997}$), 	$^{2}$ Using ($\pi_{Hip2007}$)\\
			$^{3}$ Using ($\pi_{Gaia2018}$).
\end{table}

\textbf{\,HIP 84425}; the Preliminary orbital parameters of the system were calculated by \cite{soderhjelm1999visual} then modified two times  by \cite{2010RMxAA..46..263R} and \cite{2015AJ....150...50T}.

\begin{table}[ht]
    \centering
    \caption{Modified orbital elements,  mass sums, and quality controls of the system HIP 84425, along with previous ones of: (1) \cite{soderhjelm1999visual}, (2)   \cite{2010RMxAA..46..263R} , and (3) \cite{2015AJ....150...50T}}
    \label{tab44}
    \begin{tabular}{cccccc}
    \noalign{\smallskip}
			\hline\hline
			\noalign{\smallskip}
			&	&\multicolumn{4}{c}{System HIP 84425} \\
			\cline{3-6}
			\noalign{\smallskip}
			Parameter	& Units& (1)& (2) &(3) &This work\\
			\hline
			$ P$ {$\pm$ $\sigma_{\rm P}$} & [yr]  &$23.3$      &$14.215\pm 0.050$ & $14.23\pm 0.03$ & $14.247\pm 0.057$\\
			
			$ T_0$ {$\pm$ $\sigma_{\rm T_0}$} & [yr]&$1986.5$  & $1985.98 \pm 0.17$  & $1985.97 \pm 0.06$ & $ 1985.98\pm 0.15$\\
			
			$ e$ { $\pm$  $\sigma_{\rm e}$}  & -  &$0.5$ &$0.491\pm 0.005 $& $0.476 \pm 0.01$ & $0.465\pm 0.008$\\
			
			$ a $ { $\pm$ $\sigma_{\rm a}$}& [arcsec]&$0.28$  & $0.253\pm 0.004$& $0.249\pm 0.02$ & $0.253\pm0.002$\\
			
			$ i $ { $\pm$ $\sigma_{\rm i}$} & [deg]&$123$  &$115.2 \pm 1.1$  & $ 115.3\pm 0.6$ &  $115.05\pm 0.75$\\
			
			$\Omega$ {$\pm$ $\sigma_{\rm \Omega}$}  & [deg]&$4$    &  $190.41\pm 0.62$ & $191.81\pm 0.47$
			&$192.06\pm0.37 $\\
			
			$ \omega$ {$\pm$ $\sigma_{\rm \omega}$}	& [deg]&$349$  &$135.2\pm 2.5$& $137.2 \pm 1.0$&  $137.24\pm 2.04$ \\
			
			$\rm M_{dyn} {\pm\sigma_{\rm M}}^{1}$ & [$M_\odot$] &1.17 &$2.305\pm0.379$  &$2.201\pm0.632$ & $2.307\pm0.364$
			\\
			
			$\rm M_{dyn} {\pm\sigma_{\rm M}}^{2}$ & [$M_\odot$] &1.16 &$2.286\pm0.171$ &$2.183\pm0.539$ & $2.287\pm0.133$\\
			
			$\rm M_{dyn}{\pm\sigma_{\rm M}}^{3}$ & [M$_\odot$]  &1.45 & $2.841\pm0.197$  & $2.713\pm0.666$  & $2.842\pm0.146$\\
			
						\hline
						$rms (\theta)$ &[deg]& &0.2 & 1.26 & 0.06\\
						$rms$ ($\rho$)&[arcsec]&  & 1.0&0.93&0.0001\\
			\hline\hline
			\noalign{\smallskip}
		\end{tabular}\\
			Notes.\\
		$^{1}$ Using ($\pi_{Hip1997}$), 	$^{2}$ Using ($\pi_{Hip2007}$)\\
			$^{3}$ Using ($\pi_{Gaia2018}$).
\end{table}

In this work, we use 16 points of the relative positional measurements ($\rho$) and ($\theta$) from the year 1959 to the year 2015, the latest point was published by \cite{2016AJ....151..153T}, and is used for first time in this work.

We calculate the total dynamical mass  using  Kepler's third law, where we use the semi-major axis and the period from the orbital solution and  the parallax measurements from either Hipparcos \cite{1997yCat.1239....0E}, Hipparcos 2007 \cite{2007A&A...474..653V}, or Gaia DR2 \cite{gaia2018vizier}.

 Kepler's third law for binary stars can be expressed as follows:

\begin{eqnarray}
\label{eq31}
\rm M_{dyn} = \rm M_A + \rm M_B = (\frac{\textit{a}^3}{\pi^3 P^2 } ) \rm M_\odot
\end{eqnarray}
and the formal error is given by

\begin{eqnarray}
\label{eq32}
\frac{\sigma_{\rm M} }{\rm M} = \sqrt{9(\frac{\sigma_\pi}{\pi})^2+9(\frac{\sigma_a}{a})^2+4(\frac{\sigma_P}{P})^2}.
\end{eqnarray}
\subsection{Physical parameters}\label{22}

The first step in applying the analysis using Al-Wardat's method is to  determine correctly the magnitude difference between the two components . For the system  HIP 19206, we find it $\triangle m=1.4\pm0.02$ mag, which is the average  measurements  under the V-band filters (541-551) nm (see Table~4). While for HIP 84425 we find it $\triangle m=1.6\pm0.16$ mag as the average value of the   V-band  filters  (543-551) nm measurements listed in  Table~4.

\begin{table}[ht]
    \centering
    \caption{Magnitude difference between the components of both systems;	HIP 19206 and HIP 84425. We listed here only the values taken under filters close to the V-band filter.}
			\label{tab55}
			 \begin{tabular}{c|cccc}
			 \noalign{\smallskip}
			\hline\hline
			%	\noalign{\smallskip}
		HIP &	$\triangle m$& ${\sigma_{\triangle m}}$& Filter ($\lambda/\Delta\lambda$)&  Reference  \\
		&	(mag) & & ($~\rm nm$)&    \\
			\hline
			%	\noalign{\smallskip}
			HIP 19206&
			$1.39$ &   0.02  & $545/20 $& \cite{2007AstBu..62..339B} \\				
			&$1.40$ &   $*$  & $541/88 $& \cite{2008AJ....136..312H}\\		
			&$1.43$ &     $*$  &$550/40 $& -- \\
			&$1.45$ &   $*$   & $550/40$ & -- \\		
			&$1.48$ &    $*$  &$550/40 $ & -- \\
			&$1.4$ &  $*$     & $551/22 $  &  \cite{2010AJ....139..743T} \\
			&$1.5$ &   $*$   &$543/22 $& \cite{2016AJ....151..153T}\\			
			
			\hline
			HIP 84425&
			$1.6$ &    $*$   & $551/22 $& \cite{2010AJ....139..743T} \\				
			&$1.6$ &    $*$   & $543/22 $& \cite{2015AJ....150...50T}   \\		
			&$1.7$ &      $*$  &$543/22$ & \cite{2016AJ....151..153T}       \\	
			\hline\hline
    \end{tabular}
    \label{tab:my_label}
\end{table}

Using the values of $\triangle m$ along with the apparent visual magnitudes of the systems ($m_{v(HIP19206)}=6.88$ mag, $m_{v(HIP 84425}=5.95$ mag )
(see Table~1), we calculate the apparent visual magnitudes of individual components using Equs.~3 and 4. The results are: $ m_v^A=7^{m}.14\pm0.01 $ and $ m_v^B=8^{m}.54\pm0.02$ for HIP 19206, and: $ m_v^A=6^{m}.17\pm0.03 $ and $ m_v^B=7^{m}.77\pm0.16$ for HIP 84425.

\begin{eqnarray}
\label{eq111}
m_v^A=m_v+2.5\log(1+10^{-0.4\triangle m}),
\end{eqnarray}

\begin{eqnarray}
\centering
\label{eq222}
m_v^B=m_v^A+{\triangle m}.
\end{eqnarray}

Where we used the following relations to calculate the error values of the apparent visual magnitudes of the individual components of the system:

\begin{eqnarray}
\label{eq333}
\sigma^2_{m^A_{v}} = {\sigma_{ m_{v}}^2+(\frac{1}{1+10^{+0.4\triangle m}})^2\sigma_{\triangle m}^2},
\end{eqnarray}

\begin{eqnarray}
\label{eq3311}
\sigma^2_{m^B_{v}} = {\sigma_{ m^A_{v}}^2+\sigma_{\triangle m}^2}
\end{eqnarray}

%The results of the apparent visual magnitudes associated with the  parallax from Gaia (Collaboration et al. 2018) ($\pi_{HIP19206}=23.4039\pm0.4727$,$d=42.7279$), ($\pi_{HIP84425}=30.4039\pm0.4728$,$d=32.8526$).
As a result, it is found that the absolute visual  magnitudes for the primary and secondary components of the systems  HIP 19206 are: $ M_V^A=3^{ m}.96\pm0.04$ and $ M_V^B=5^{ m}.36\pm0.05$ respectively, and those for the system HIP 84425 are: $M_V
^A = 3^{m}.58\pm0.05 $, $M_V
^B = 5^{m}.17\pm0.03 $, using the following equation (Heintz, 1978):

\begin{eqnarray}
\label{eq3}
M_V=m_v+5-5\log(d)-A_V,
\end{eqnarray}
Here, the interstellar extinction, $\rm A_V$ of HIP 19206 is neglected because this system is a nearby one. %However, the interstellar measurement for Hip 84425 is  27 as given in Table~1, which is a strange value and  needs more investigation, that is why it is neglected  in our calculations.

The errors of the absolute visual magnitudes of the components A and B of the system are calculated by using the following equation:

\begin{eqnarray}
\label{eq341}
\sigma^2_{ M^{*}_{V}} ={\sigma_{ m^{*}_{v}}^2+(\frac{ \log\rm e}{0.2\pi})^2\sigma_{\pi}^2} ~;~~~~~ \textit{*}=A,B.
\end{eqnarray}

\noindent where  $ \sigma_{m^*_{v}} $ are the errors of the apparent magnitudes of the A and B components expressed in Equations~5~\&~6.

Based on the estimated preliminary absolute magnitudes ($\rm M_V$) of the individual components of the two systems, we can find the preliminary values of the effective temperature and the bolometric correction for each component as taken from the Tables of \cite{2005oasp.book.....G} and \cite{lang1992astrophysical} . For HIP\,19206 $T^A_{\rm eff}=6160\,\rm K$ ,$T^B_{\rm eff}=5550\,\rm K$  and $(B.C)_{\rm A}=-0.12$, $(B.C)_{\rm B}=-0.30$, and for HIP 84425 $T^A_{\rm eff}=6520\,\rm K$ ,$T^B_{\rm eff}=5650\,\rm K$  and $(B.C)_{\rm A}=-0.12$, $(B.C)_{\rm B}=-0.21$, for the primary and secondary components of the two systems, respectively.

Furthermore, we use Equs.~9,10,11\& 12 to calculate the input parameters and to double check them after getting the best fit between the synthetic and observed photometry.

% calculate the preliminary values of the bolometric magnitude, luminosity,radius and gravity acceleration for each component of the systems.

%For HIP 19206 as: $M^A_{\rm bol}=3.80$, $M^B_{\rm bol}=5.06$, $L_{\rm A}=2.33{L_\odot}$, $L_{\rm B}=0.73{L_\odot}$, $R_{\rm A}=1.34{R_\odot}$, $R_{\rm B}=0.92{R_\odot}$, $\rm log g_{\rm A}=4.255$, and $\rm log g_{\rm B}=4.40$.

%For HIP 84425 as: $M^A_{\rm bol}=3.46$, $M^B_{\rm bol}=4.97$, $L_{\rm A}=3.16{L_\odot}$, $L_{\rm B}=0.73{L_\odot}$, $R_{\rm A}=1.4{R_\odot}$, $R_{\rm B}=0.93{R_\odot}$, $\rm log g_{\rm A}=4.28$, and $\rm log g_{\rm B}=4.43$.

\begin{eqnarray}
M_{bol}=M_V+B.C,
\end{eqnarray}

\begin{eqnarray}
\label{eq81}
\log\frac{L}{L_\odot}= \frac{M_{bol}^\odot-M_{bol}}{2.5},
\end{eqnarray}

\begin{eqnarray}
\label{eq81}
\log\frac{R}{R_\odot}= 0.5log\frac{L}{L_\odot}-2log\frac{T}{T_\odot} ,
\end{eqnarray}

\begin{eqnarray}
\label{eq82}
\log g = \log\frac{\rm M}{\rm M_\odot}- 2\log\frac{R}{R_\odot} + 4.43.
\end{eqnarray}

\noindent where $T_\odot=5777\,\rm K$,$R_\odot=6.69*10^{\rm 8} \rm m$ and $M_{bol}^\odot=4^{\rm m}.75$.

In order to obtain the best stellar parameters, as we mentioned above, we need to build a synthetic SED of the system based on the input parameters and on grids of line-blanketed model atmospheres (ATLAS9) \cite{1994KurCD..19.....K}. This parameter could not be built unless we are fully aware of the information about its distance d and  radius R (see Figures 4 \& 5). Hence, the entire synthetic SED as it would be appears at the Earth surface for the entire binary system  is related to the individual synthetic SEDs of the components according to the following equation \cite{2002BSAO...53...51A, 2012PASA...29..523A}:

\begin{eqnarray}
\label{eq66}
F_\lambda \cdot d^2 = H_\lambda ^A \cdot R_{A} ^2 + H_\lambda ^B
\cdot R_{B} ^2,
\end{eqnarray}
\noindent

\noindent

where $F_\lambda$ is the flux for the entire synthetic  SED of the entire binary system at the Earth, $H_\lambda ^A $ and  $H_\lambda ^B$ are the fluxes of the primary and secondary  components, in units of ergs cm$^{-2}$\rm s$^{-1}$ \AA$^{-1}$, while $ R_{A}$ and $ R_{B}$ are the radii of the primary and secondary components in solar units.

\begin{figure}

  \begin{subfigure}{7cm}
    \centering\includegraphics[width=7cm]{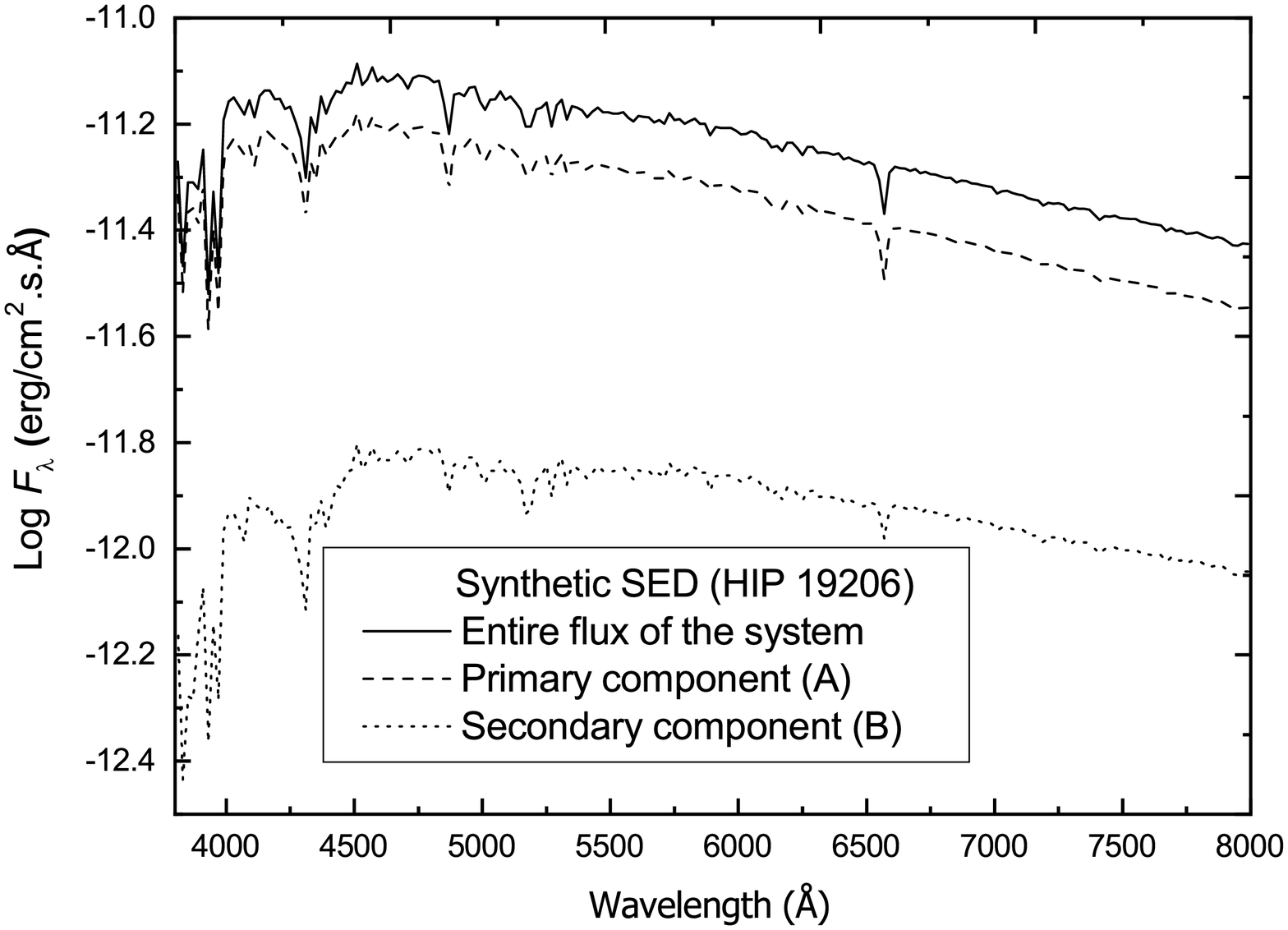}
    \caption{HIP 19206}
  \end{subfigure}
    \begin{subfigure}{7cm}
    \centering\includegraphics[width=7cm]{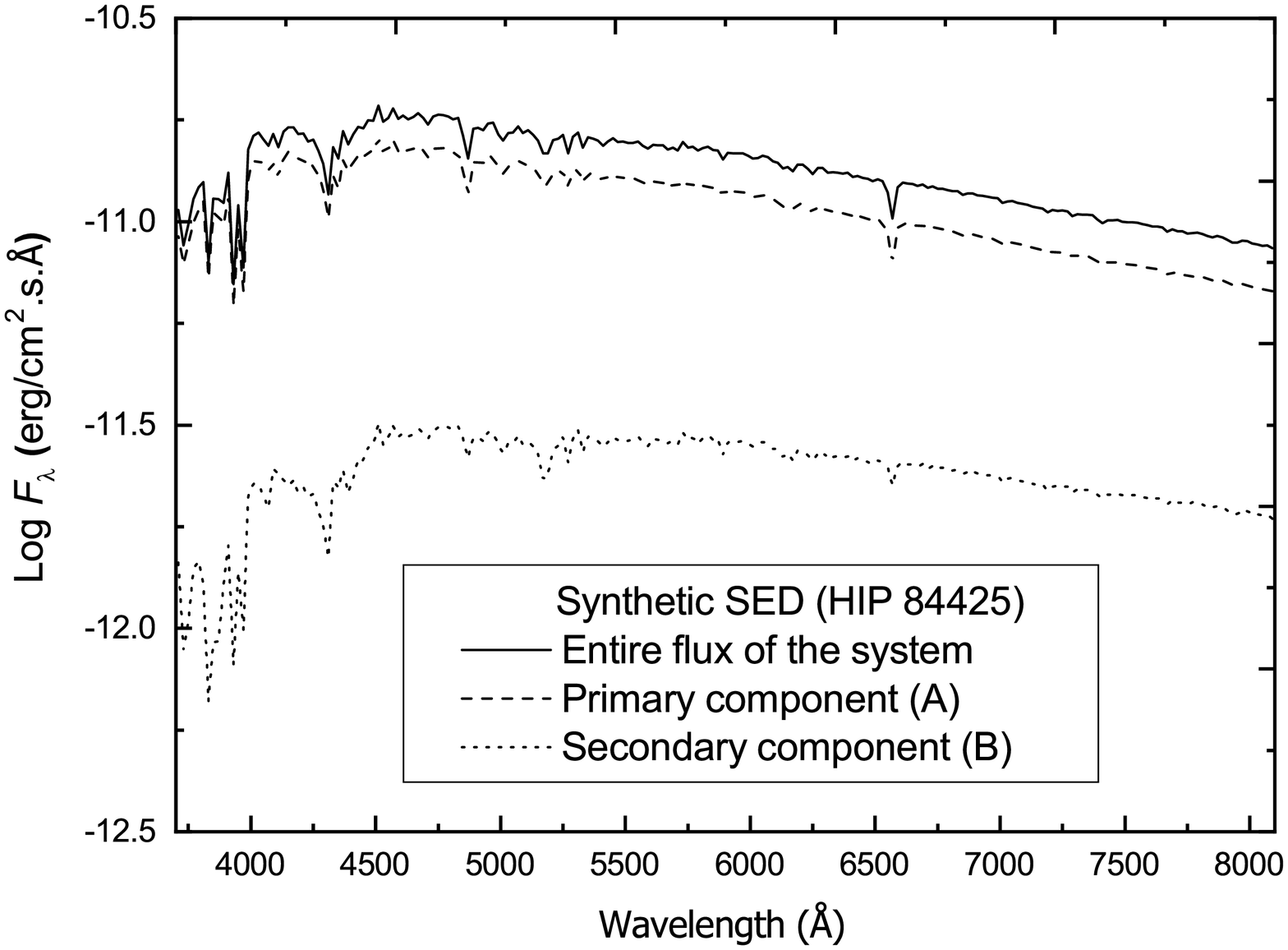}
    \caption{HIP 84425}
  \end{subfigure}

\caption{The entire flux and individual synthetic SEDs of the binary system HIP 19206 and HIP 84425 using Kurucz blanketed models \cite{1994KurCD..19.....K} (ATLAS9).}
    \label{sed}
\end{figure}

Since we do not have  observational  spectral energy distribution  for the systems,  we depend on the magnitudes and colour indices  as a reference for the best fit with the synthetic ones, which ensures the reliability of the estimated  parameters. This technique is a part of Al-Wardat's complex method for analyzing CVBSs.

\section{Synthetic photometry}
Stellar parameters depend mainly on the best fit of the magnitudes and color indices between the entire observational and synthetic SED of the system. Therefore, we calculate the individual and total synthetic magnitudes and colour indices of the  systems by integrating the total fluxes over each bandpass of the named photometrical system divided by that of the reference star  (Vega),  using the following equation \cite{2012PASA...29..523A}:

\begin{equation}\label{55}
m_p[F_{\lambda,s}(\lambda)] = -2.5 \log \frac{\int P_{p}(\lambda)F_{\lambda,s}(\lambda)\lambda{\rm d}\lambda}{\int P_{p}(\lambda)F_{\lambda,r}(\lambda)\lambda{\rm d}\lambda}+ {\rm ZP}_p\,
\end{equation}

\noindent	where $m_p$ is the synthetic magnitude of the passband $p$, $P_p(\lambda)$ is the dimensionless sensitivity function of the passband $p$, $F_{\lambda,s}(\lambda)$ is the synthetic SED of the object and $F_{\lambda,r}(\lambda)$ is the SED of the reference star (Vega).  Zero points (ZP$_p$) from \cite{2007ASPC..364..227M} are used.

\begin{table}[ht]
\small
	\begin{center}
		\caption{ Magnitudes and colour indices  of the entire synthetic spectrum and individual components of  HIP 19206 and HIP 84425.}
		\label{tab66}
		\begin{tabular}{cc|ccc|ccc}
			\noalign{\smallskip}
			\hline\hline
			\noalign{\smallskip}
			Sys. & Filter
		&	\multicolumn{3}{c}{HIP 19206}&  \multicolumn{3}{c}{HIP 84425}\\
				\cline{3-5}
			\cline{6-8}
		
				\noalign{\smallskip}

			&     &  Entire Synth. &   A    &     B    &  Entire Synth.&   A    &     B  \\
			&&$\sigma=\pm0.03$&&&$\sigma=\pm0.03$\\
			\hline
			\noalign{\smallskip}
			Joh-          & $U$ & 7.55 & 7.73 & 9.62& 6.63 & 6.77 & 8.95 \\
			Cou.          & $B$ & 7.45   &  7.67 &  9.33& 6.53   &  6.71 &  8.58  \\
			& $V$ & 6.88 &  7.14 &  8.58& 5.95 &  6.17 &  7.79 \\
			& $R$ & 6.56  &6.84   &8.18  &5.63   & 5.87&7.37   \\
			&$U-B$ &  0.09 &0.05  &0.29 & 0.10  & 0.06 & 0.37\\
			&$B-V$ &  0.58 &  0.53 & 0.75 & 0.58  &  0.54 &  0.79 \\
			&$V-R$&  0.32 &  0.93 &0.40 & 0.32  &  0.93 & 0.40 \\
			\hline
			\noalign{\smallskip}
			Str\"{o}m.    & $u$ &8.72  & 8.89 &10.76  &7.80  & 7.94 &  10.10 \\
			& $v$ & 7.78 & 7.97 & 9.73 & 6.86 & 7.02  & 9.01 \\
			& $b$ & 7.21 & 7.44 & 8.98 & 6.28 & 6.48 &  8.22 \\
			&  $y$& 6.86 &7.11  & 8.54 & 5.92 & 6.14 & 7.75 \\
			&$u-v$& 0.94 & 0.92 & 1.04& 0.95 & 0.93 & 1.08 \\
			&$v-b$& 0.56 & 0.53 & 0.74& 0.58 & 0.54 & 0.79 \\
			&$b-y$&  0.35 & 0.33& 0.44& 0.36 & 0.34& 0.47 \\
			\hline
			\noalign{\smallskip}
			Tycho       &$B_T$  & 7.59   & 7.79 & 0.86& 6.67   & 6.84 & 8.79  \\
			&$V_T$  & 6.95   & 7.20& 8.66 & 6.02   & 6.23& 7.87 \\
			&$B_T-V_T$& 0.65 & 0.59 & 0.86& 0.65 & 0.60 & 0.92\\
			\hline\hline
			\noalign{\smallskip}
		\end{tabular}
	\end{center}

\end{table}

\begin{table}[!ht]
	\small
	\begin{center}
		\caption{Comparison between the observational and synthetic
			magnitudes and colours indices for both systems.} \label{tab77}
		\begin{tabular}{c|ccc}
			\noalign{\smallskip}
			\hline\hline
			\noalign{\smallskip}
		
			\noalign{\smallskip}
		HIP&	Filter	& Observed  & Synthetic (This work)\\
			&& ($\rm mag$) & ($\rm mag$) \\
			
			\hline
			\noalign{\smallskip}
				HIP 19206&$V_{J}$ & $6.88$ & $6.88\pm0.03$\\
			
			&$B_T$ & $7.566\pm0.008$   &7.59$\pm0.03$\\
			&$V_T$  & $6.957\pm0.006$   &$6.95\pm0.03$\\
			&$(B-V)_{J}$&$ 0.576\pm0.004$ &$ 0.58\pm0.03$\\
			
		&	$\triangle m$  & $ 1.4$  & $ 1.4$\\
				\hline
							\noalign{\smallskip}
				HIP 84425&$V_{J}$ &  $5.95$ & $5.95\pm0.03$\\
			
		&	$B_T$ &  $6.67\pm0.013$   &6.67$\pm0.03$\\
			&$V_T$  & $6.015\pm0.006$   &$6.02\pm0.03$\\
			&$(B-V)_{J}$&$ 0.58\pm0.002$ &$ 0.58\pm0.03$\\
			
		&	$\triangle m$  & 1.6  & 1.6\\
			\hline\hline \noalign{\smallskip}
		\end{tabular}\\
		
	\end{center}
\end{table}

%\begin{figure}[ht]
%	\centering
%	\includegraphics[angle=0,width=12cm]{Fig3.pdf}
%	\caption{The entire flux and individual synthetic SEDs of the binary system HIP 19206 using Kurucz blanketed models (Kurucz 1994) (ATLAS9).\label{fig03}}
%\end{figure}
%\begin{figure}[ht]
%	\centering
%	\includegraphics[angle=0,width=12cm]{Fig4.pdf}
%	\caption{The entire flux and individual synthetic SEDs of the binary system HIP 84425 using Kurucz blanketed models (Kurucz 1994) (ATLAS9).}
%	\label{fig02}
%\end{figure}

We used the best fit  between the observational and synthetic entire magnitudes, magnitude differences of the components ($\triangle m$ = $V^{B}_{J}$-$V^{A}_{J}$) and colour indices  to judge the accuracy  of the estimated parameters.

The entire flux is calculated from the individual fluxes of the components using  Equ.~13~ and depending on the parallax of the system.

As mentioned above, we get the color indices of the individual components and entire synthetic SEDs of two systems, in three photometrical systems; Johnson: $U$, $B$, $ V$, $R$, $U-B$, $B-V$, $V-R$; Str\"{o}mgren: $u$, $v$, $b$,
$y$, $u-v$, $v-b$, $b-y$ and Tycho: $B_{T}$, $ V_{T}$, $B_{T}-V_{T}$, which are listed in Table~5.

\section {Results and discussion}\label{4}

%The results of the new orbital solutions show that the RMS values of $\triangle\rm \theta$ and  $\triangle\rm \rho$ for HIP 19206 are 0.\degr10 and 0.''0002, while for HIP 84425 are 0.\degr06 and 0.''0001. These values are better than those given in previous solutions, which raise the accuracy of the solutions.

%Tables ~3 and ~4 show the results of the accurate orbital parameters of the close binary systems HIP 19206 and HIP 84425, which are shown in Figures ~1 \& ~2.

Table~5 lists the results of the synthetic magnitudes and colour indices of the entire systems and individual components  under three photometric systems: Johnson: $U$, $B$, $ V$, $R$, $U-B$, $B-V$, $V-R$; Str\"{o}mgren: $u$, $v$, $b$,
$y$, $u-v$, $v-b$, $b-y$ and Tycho: $B_{T}$, $ V_{T}$, $B_{T}-V_{T}$).
And Table ~6 gives a comparison  between the entire synthetic magnitudes and colour indices of the two systems with the available observed ones within three photometric systems.
The consistency between the synthetic and observed photometry indicates the reliability of the fundamental parameters of the two systems.

Hereafter we discuss the results of each system separately:

%To estimate the stellar masses of the systems and the positions based on the H-R diagram, our calculations for the input physics have carried out with the stellar evolution tracks developed by Girardi et al. (2000b). The evolutionary tracks have shown that the two systems belong to the last stage of the main sequence (see Figure~4). According to that, we calculate the stellar masses for both systems using two different methods Al-Wardat's and dynamical methods.

\textbf{HIP 19206}:
Table~2 lists the modified orbital elements as per the new orbital solution, which used  a total of 41 relative positional measurements,  along with the elements of the previous solutions of  \cite{2006BSAO...59...20B},  \cite{2015AJ....150...50T} and  \cite{2019IAUDS.199....1C}. The  RMS values of the new  solution $\triangle\rm \theta = 0.10\degr$ and  $\triangle\rm \rho = 0.0002\arcsec$      are better than those of the previous solutions.

Table~7 lists the final fundamental parameters of the system. It shows that the  dynamical mass sum of the system using  Gaia parallax ($23.4039\pm 0.4727$) is given by $2.087\pm 0.128$ which is the closest to the mass sum given by Al-Wardat's method as $2.14\pm 0.15$. Which means that Gaia parallax measurement is the best among the other measurements, but in order to achieve the best consistency between the dynamical mass sum and that of Al-Wardat's method, we propose a new dynamical parallax for the system as $22.97\pm 0.95$ mas.

It is worthwhile to mention here that this work was done before the release of Gaia DR3 in December 2020 which gives a very close trigonometric parallax for the system as $22.3689\pm 0.4056$ mas \cite{brown2020gaia}.

The reason why we introduced a new parallax measurement is that parallax measurements of binary and multiple systems are, in some cases, distorted by the orbital motion of the components of such systems as noted by other several researchers (i.e. \cite{1998AstL...24..673S}).

The fundamental parameters of  the systems' components, their positions on the evolutionary tracks of \cite{2000A&AS..141..371G} (Fig.~4a), and on the isochrones of \cite{2000yCat..41410371G} (Fig~5a), show that they are a twin  solar type main sequence stars with a metalicity of 0.019. Which predicts that fragmentation is the most probable formation process for the system.

\textbf{HIP 84425}:
Table~3 lists the modified orbital elements as per the new orbital solution, which used  a total of 13 relative positional measurements,  along with the elements of the previous solutions of \cite{soderhjelm1999visual},    \cite{2010RMxAA..46..263R} , and \cite{2015AJ....150...50T}. The  RMS values of the new  solution $\triangle\rm \theta = 0.06\degr$ and  $\triangle\rm \rho = 0.0001\arcsec$      are much better than those of the previous solutions.

Table~8 lists the fundamental parameters of the system as given by Al-Wardat's method using two parallax measurements; Hipparcos 2007 and Gaia 2018. So, we have to decide which set of parameters represent better the system.

In order to do so, it should be  remembered that the dynamical mass sum is  highly affected by the parallax,  while the masses estimated using  Al-Wardat's method are not as highly  affected  by the change of the parallax. This is clear in Fig.~4b which shows the positions of the components of HIP 84425 on the evolutionary tracks when using both the parallax given by Hipparcos 2007 and that given by Gaia 2018.

Now, depending on the mass sum of Al-Wardat's method as $2.2\pm0.15$ mas, we estimate a new dynamical parallax as ($33.26\pm1.5$), which is closer to that of Hipparcos 2007 as ($32.69\pm0.59$). So, we adopt the fundamental parameters given in Table~8 columns 3 and 4.

The fundamental parameters of  the systems' components, their positions on the evolutionary tracks of \cite{2000A&AS..141..371G} (Fig.~4b), and on the isochrones of \cite{2000yCat..41410371G} (Fig~5b), show that they are a twin  solar type in their final main sequence stage with a metalicity of 0.019. Which predicts that fragmentation is the most probable formation process for the system.

\begin{table}
\centering
	\small
	\begin{center}
		\caption{The fundamental parameters of the individual components of  HIP 19206 as estimated using Al-Wardat's Method.}
		\label{tab88}
		\begin{tabular}{cccc}
			\noalign{\smallskip}
			\hline\hline
			&
			&\multicolumn{2}{c}{HIP 19206}

			\\
			
			\cline{3-4}

		\noalign{\smallskip}
			Parameters & Units	& HIP 19206 A & HIP 19206 B \\
			\hline
			\noalign{\smallskip}
			
			$ T_{ eff}$ {$\pm$ $\sigma_{ T_{eff}}$}& [~K~] & $6300\pm100$ & $5600\pm100$  \\
			
			 {$R\pm \sigma_{R}$}  & [R$_{\odot}$] &$1.417\pm0.06$&$0.978\pm0.05$\\
			
			$\log\rm g$ {$\pm$ $\sigma_{\rm log g}$} & [$cm/s^2$] & $4.18\pm0.11$ &$4.40\pm0.13$\\
			
			$ L $ {$\pm$ $\sigma_{\rm L}$} & [$\rm L_\odot$] &$2.842\pm0.30$ & $0.845\pm0.10 $\\
			
			$ M_{bol}$ {$\pm$ $\sigma_{\rm M_{bol}}$} & [mag] & $3.81\pm0.08$ &$4.87\pm0.08$ \\
			$ M_{V}$ {$\pm$ $\sigma_{\rm M_{V}}$} & [mag] & $3.97\pm0.13$ & $5.08\pm0.14$\\
			Sp. Type$^{1}$ & &  F7V & G6V \\
		
			$\rm M$	& [$\rm M_{\odot}$] &$1.23\pm0.15$&$0.91\pm0.10$\\

			\hline
				$\rm M_{tot}$	&[$\rm M_{\odot}$]&  \multicolumn{2}{c}{$2.14 \pm 0.15 $}	\\
				$\rm M_{dyn}^{*} {\pm\sigma_{\rm M}}$ & [$M_\odot$] & \multicolumn{2}{c}{$1.873\pm0.0.215$} \\
				$\rm M_{dyn}^{**} {\pm\sigma_{\rm M}}$ & [$M_\odot$] & \multicolumn{2}{c}{$1.76\pm0.133$} \\
				$\rm M_{dyn}^{***} {\pm\sigma_{\rm M}}$ & [$M_\odot$] & \multicolumn{2}{c}{$2.087\pm0.128$} \\
			$\pi_{dyn}^{2}$	& [mas] &  \multicolumn{2}{c}{$22.97 \pm 0.95$}	\\				
			Age$^{2}$	& [Gyr] & \multicolumn{2}{c}{ $7\pm3$}\\
			\hline\hline
			\noalign{\smallskip}
		\end{tabular}\\
		
	{
	%Depending on the evolutionary tracks (Girardi et al. 2000b).\\
		$^{1}${Using the tables of \cite{1992adps.book.....L} and \cite{2005oasp.book.....G}}.\\
		$^{2}$Depending on the the isochrones for low- and intermediate-mass stars of
			different metallicities of~ (\cite{girardi2000evolutionary}) (see Figures ~5}).\\
				$^{*}$ Using ($\pi_{Hip1997}$), 	$^{**}$ Using ($\pi_{Hip2007}$),
			$^{***}$ Using ($\pi_{Gaia2018}$).
			
	\end{center}

\end{table}

\begin{table}
\centering
	\small
	\begin{center}
		\caption{The fundamental parameters of the individual components of HIP 84425 as estimated using Al-Wardat's Method.}
		\label{tab88}
		\begin{tabular}{ccccccc}
			\noalign{\smallskip}
			\hline\hline
			&

			&\multicolumn{4}{c}{HIP 84425}
			\\
			\cline{3-6}
			\\
		& & \multicolumn{2}{c}{HIP 2007}&\multicolumn{2}{c}{Gaia 2018}\\
			
		\noalign{\smallskip}
			Parameters & Units	&  HIP 84425 A & HIP 84425 B &HIP 84425 A & HIP 84425 B\\
			\hline
			\noalign{\smallskip}
			
			$ T_{ eff}$ {$\pm$ $\sigma_{ T_{eff}}$}& [~K~] &  $6270\pm100$ & $5470\pm100$ &$6270\pm100$&$5470\pm100$ \\
			
			 {$R\pm \sigma_{R}$}  & [R$_{\odot}$] & $1.705\pm0.06$ & $1.141\pm0.05$&$1.586\pm0.06$&$1.064\pm0.05$\\
			
			$\log\rm g$ {$\pm$ $\sigma_{\rm log g}$} & [$cm/s^2$] &  $4.09\pm0.12$ & $4.26\pm0.10$& $4.137\pm 0.11$ &$4.321\pm 0.12$\\
			
			$ L $ {$\pm$ $\sigma_{\rm L}$} & [$\rm L_\odot$] & $4.04\pm0.30 $  & $1.05\pm0.10$ &$3.49\pm0.30$ & $0.91\pm 0.15$ \\
			
			$ M_{bol}$ {$\pm$ $\sigma_{\rm M_{bol}}$} & [mag] & $3.23\pm0.08$ & $4.70\pm0.09$ & $3.393\pm0.09$ & $4.853\pm0.09$ \\
			$ M_{V}$ {$\pm$ $\sigma_{\rm M_{V}}$} & [mag] &  $3.29\pm0.13$ & $4.89\pm0.13$ &$3.453\pm0.13$ &$5.043\pm0.13$ \\
			
			Sp. Type$^{1}$ &&   F7.5V & G8V&  F7.5V & G8V  \\
			$\rm M$	& [$\rm M_{\odot}$] &$1.31\pm0.13$&$0.89\pm0.12$ &$1.29\pm0.12$& $0.88\pm0.12$\\
		
			\hline
				$\rm M_{tot}$	&[$\rm M_{\odot}$]&  \multicolumn{2}{c}{$2.2 \pm 0.15 $}	&  \multicolumn{2}{c}{$2.17 \pm 0.15 $}	\\
			
				$\rm M_{dyn} {\pm\sigma_{\rm M}}$ & [$\rm M_\odot$] & \multicolumn{2}{c}{$2.287\pm0.133$}& \multicolumn{2}{c}{$2.842\pm0.146$} \\

			$\pi_{dyn}$ 	& [mas]&  	\multicolumn{4}{c}{$33.26 \pm 1.5$}\\		
					
			Age $^{2}$	& [Gyr]&  \multicolumn{2}{c}{ $7\pm$3}\\
			\hline\hline
			\noalign{\smallskip}
		\end{tabular}\\
		
	{
	%Depending on the evolutionary tracks (Girardi et al. 2000b).\\
		$^{1}${Using the tables of \cite{1992adps.book.....L} and \cite{2005oasp.book.....G}}.\\
		
		$^{2}$Depending on the the isochrones for low- and intermediate-mass stars of
			different \\metallicities of~ (\cite{girardi2000evolutionary}) (see Figures ~5}).	
	\end{center}

\end{table}

\begin{figure}

  \begin{subfigure}{7cm}
    \centering\includegraphics[width=7cm]{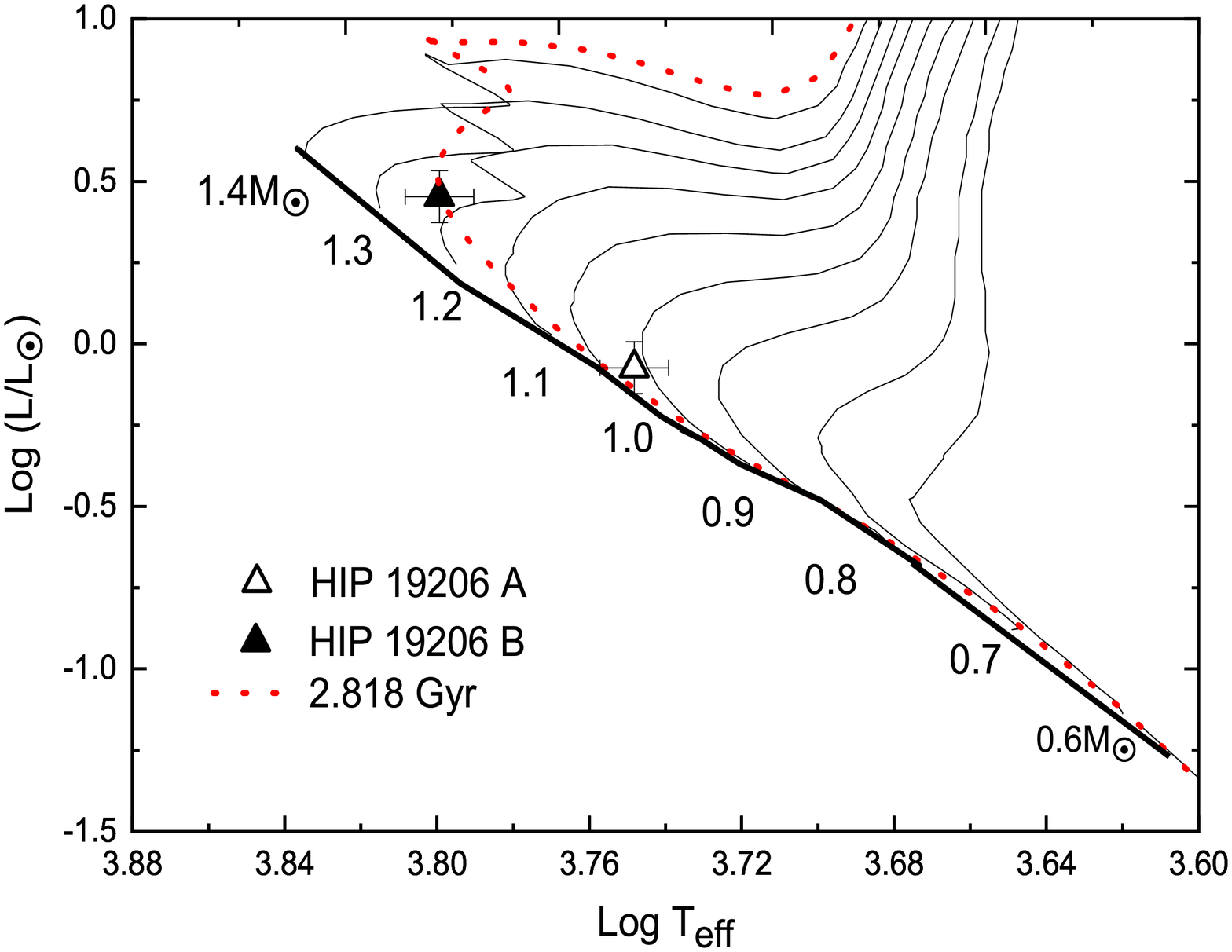}
    \caption{HIP 19206}
  \end{subfigure}
    \begin{subfigure}{7cm}
    \centering\includegraphics[width=7cm]{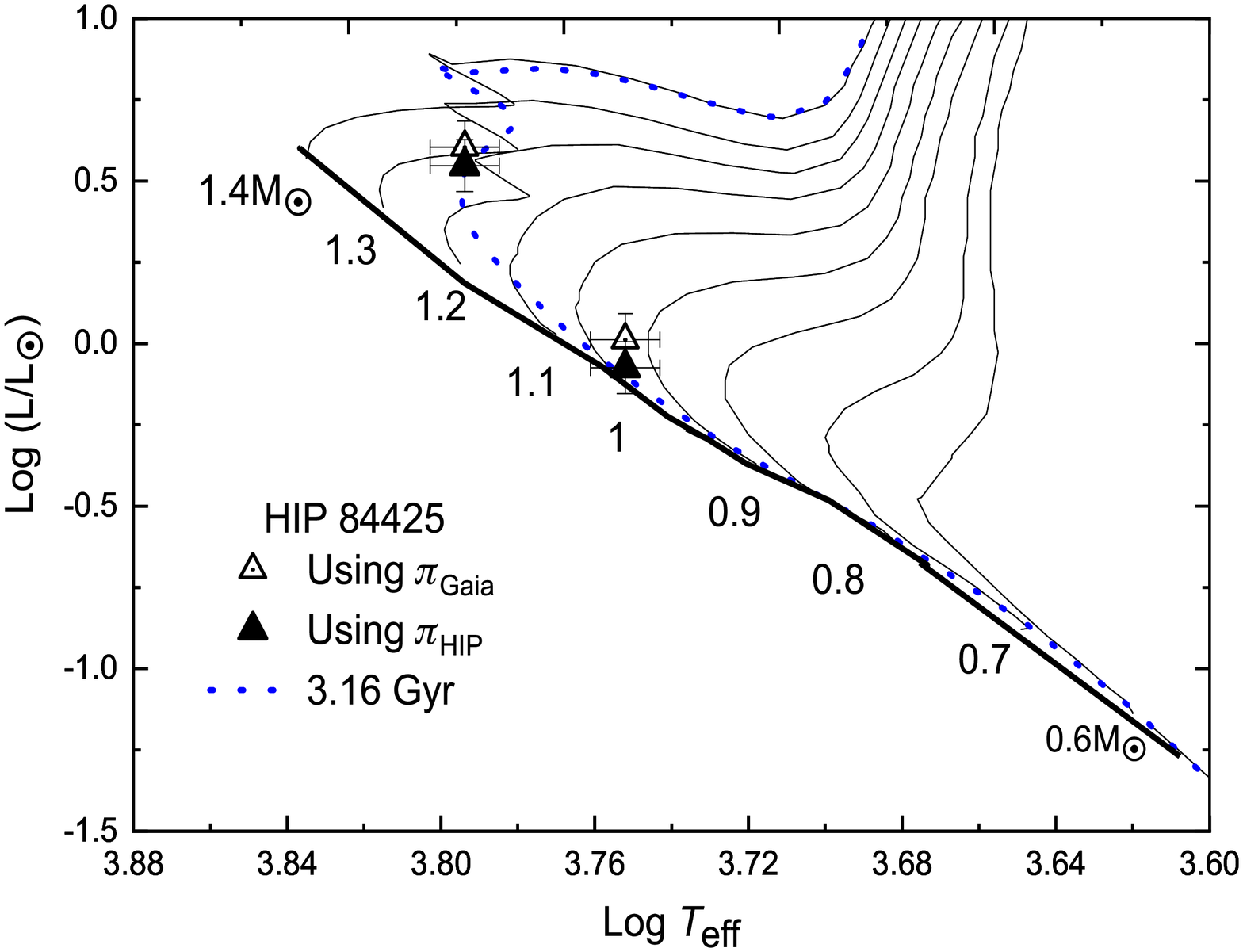}
    \caption{HIP 84425}
  \end{subfigure}

\caption{The positions of the components of HIP 19206 and HIP 84425 on the evolutionary tracks of \cite{2000A&AS..141..371G} for the masses ( 0.6, 0.7, ...., 1.4 $\rm\,M_\odot$). Figure b shows the positions of the components as been estimated using the parallaxes given by Hipparcos and Gaia (See Table~\ref{tab1}). }
    \label{evo}
\end{figure}

%\begin{figure}[ht]
%	\centering
%	\includegraphics[angle=0,width=12cm]{Fig5.pdf}
%	\caption{The evolutionary tracks of both components of HIP 19206 on the H-R diagram of masses ( 0.8, 0.9, ...., 1.3 $\rm\,M_\odot$). The evolutionary tracks were taken from~Girardi et al. (2000b). }
%	\label{evola19}
%\end{figure}
%\begin{figure}[ht]
	%\centering
	%\includegraphics[angle=0,width=14cm]{evo84425.eps}
	%\caption{The evolutionary tracks of both components of HIP 84425 on the H-R diagram of masses ( 0.8, 0.9, ...., 1.3 $\rm\,M_\odot$). The evolutionary tracks were taken from Girardi et al. (2000b). }
%	\label{evola855}
%\end{figure}

%\begin{figure}[ht]
%	\centering
%	\includegraphics[angle=0,width=12cm]{Fig7.eps}
%	\caption{The isochrones of the components of HIP 19206  on the HR diagram for low- and intermediate-mass stars of		different metallicities. The isochrones were taken from Girardi et al. (2000a). }
%	\label{Iso19}
%\end{figure}
%\begin{figure}[ht]
%	\centering
%	\includegraphics[angle=0,width=12cm]{Fig8.eps}
%	\caption{The isochrones of the components of HIP 84425  on the HR diagram for low- and intermediate-mass stars of		different metallicities. The isochrones were taken from Girardi et al. (2000b). }
%	\label{Iso844}
%\end{figure}

\begin{figure}

  \begin{subfigure}{7cm}
    \centering\includegraphics[width=7cm]{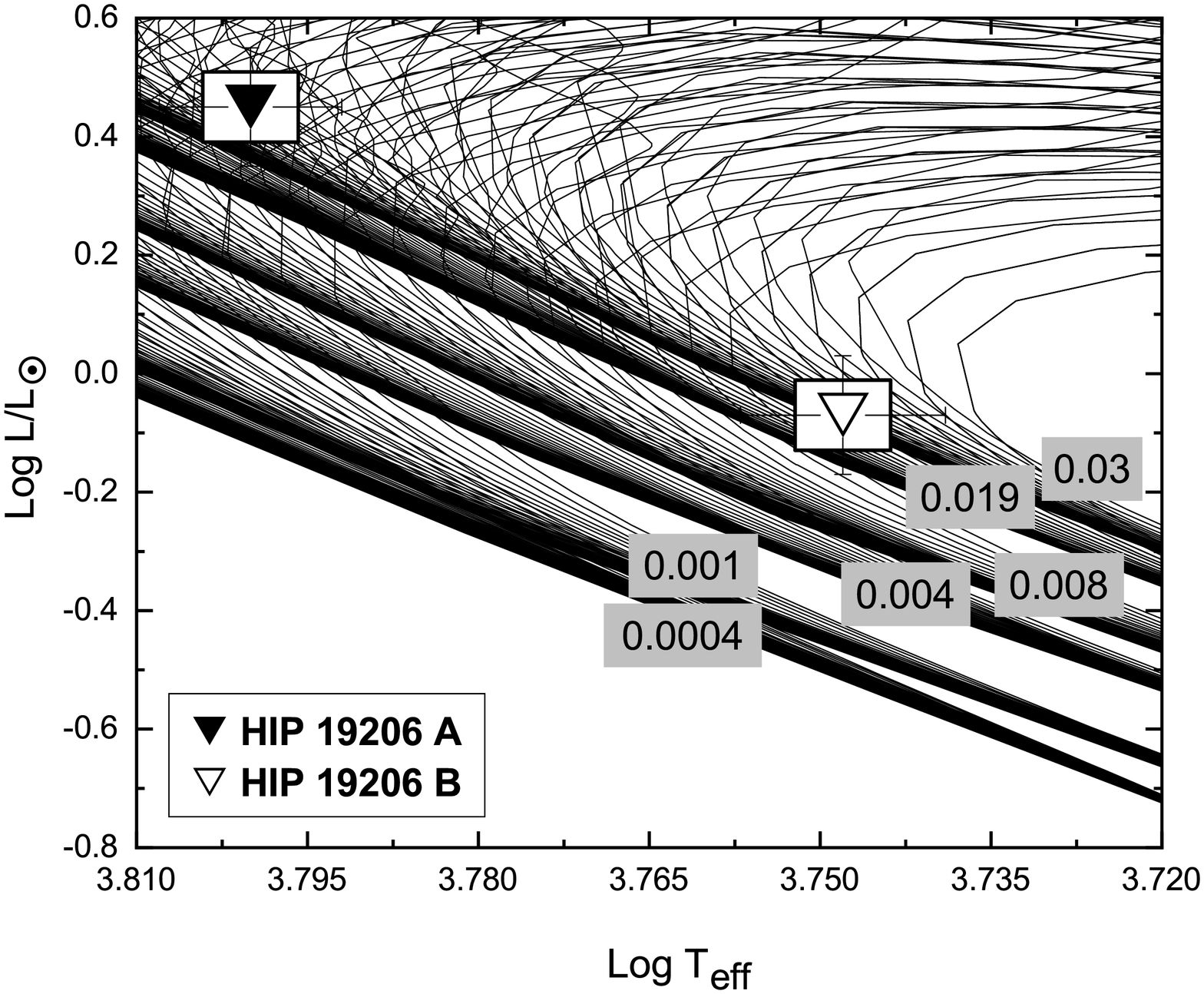}
    \caption{HIP 19206}
  \end{subfigure}
    \begin{subfigure}{7cm}
 \centering\includegraphics[width=7cm]{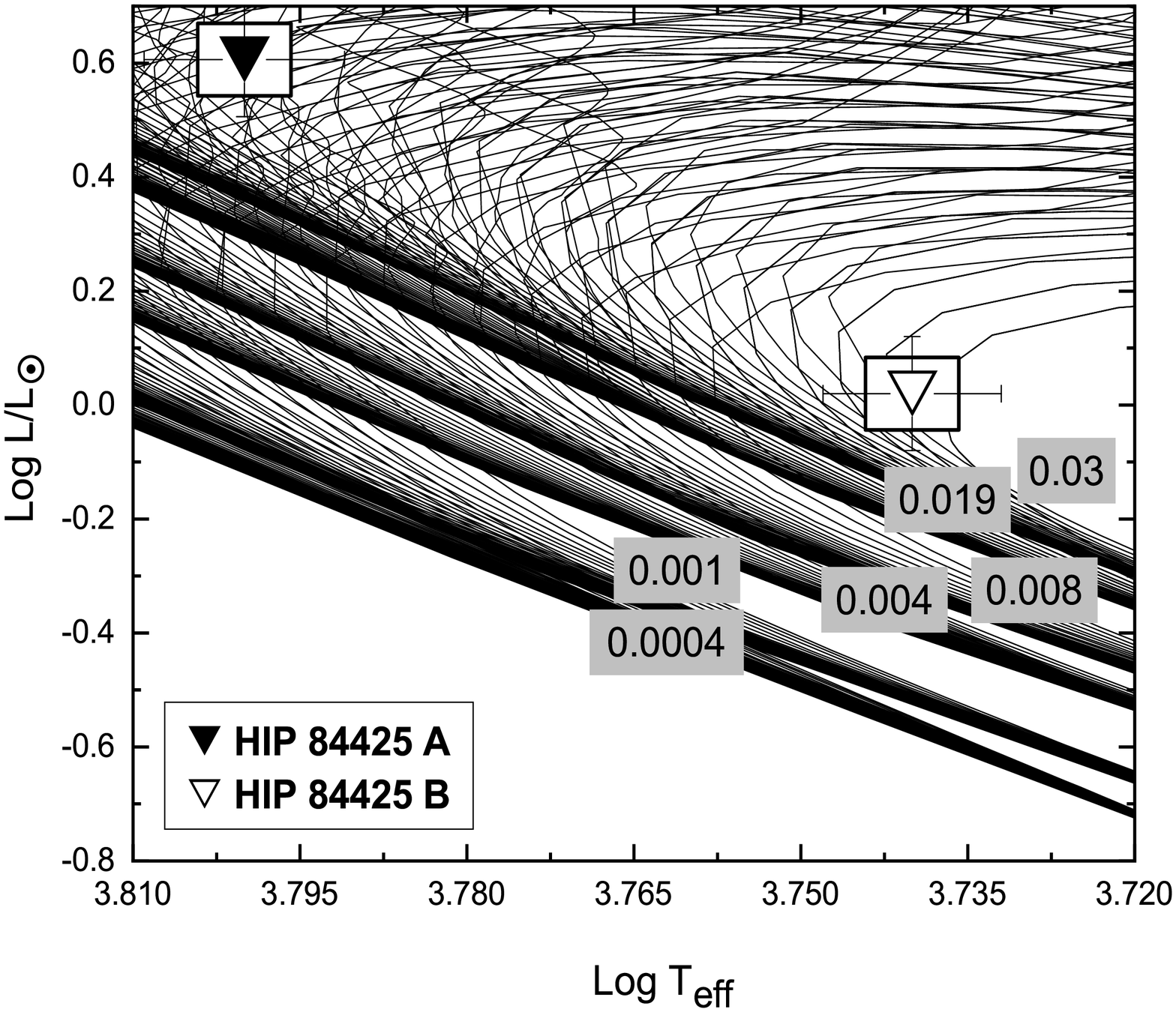}
    \caption{HIP 84425}
  \end{subfigure}

\caption{The positions of the components of HIP 19206 and HIP 84425  on the isochrones of low- and intermediate-mass stars of different metallicities. The isochrones were taken from \cite{girardi2000evolutionary}.}
    \label{iso}
\end{figure}

\section{Conclusions}
%In this contribution, we have showed that the analyzing of two CVBSs HIP 19206 and HIP 84425 systems to estimate their physical and geometrical parameters, is achieved and motivated by the Al-Wardat's method. This method is based on the accurate analysis of the synthetic spectra of binary systems associated with the Kurucz ATLAS9 line-blanketed plane parallel model atmospheres.

%The detailed analysis based on this technique, is resulted within the following set of parameters: For HIP 19206 as: $T_{\rm eff}^{A}$ =6300$\pm100 \rm K$, $T_{\rm eff}^{B} =5600\pm100 \rm K$, $\log \rm g_A = 4.35\pm0.07 $, $\log\rm g_B = 4.50\pm0.08$, $R_{A} = 1.417\pm0.05\, R_\odot$, and $R_{B}=0.978\pm0.04\, R_\odot$.
%While for HIP 84425 as: $T_{\rm eff}^{A}=6270\pm100\,\rm K$, $T_{\rm eff}^{B} =5470\pm100\,\rm K,$, $\log \rm g_A=4.30\pm0.13 $, $\log\rm g_B=4.45\pm0.11$, $R_{A}=1.705\pm0.06\, R_\odot$, and $R_{B}=1.141\pm0.04\, R_\odot$.

%For HIP\,19206 $T^A_{\rm eff}=6160\,\rm K$ ,$T^B_{\rm eff}=5550\,\rm K$  and $(B.C)_{\rm A}=-0.12$, $(B.C)_{\rm B}=-0.30$, while for HIP 84425  $T^A_{\rm eff}=6520\,\rm K$, $T^B_{\rm eff}=5650\,\rm K$ and $(B.C)_{\rm A}=-0.12$, $(B.C)_{\rm B}=-0.21$

Using the recent parallax measurements given by Gaia  (Gaia DR2), we were able to present a complete analysis for the two CVBSs HIP 19206 and HIP 84425.

The technique uses a dynamical and a spectrophotometrical methods in a complementary iterated way, to estimate the complete set of their fundamental parameters.

In order to get the precise fundamental parameters of the systems, We used a combination of a dynamical method (ORBITX) \cite{2016AJ....151..153T}, and a computational spectrophotometrical method (Al-Wardat's method for analyzing CVBMSs as , which employs grids of \cite{1994KurCD..19.....K} line-blanketed plane-parallel model atmospheres (ATLAS 9) for single stars).

As well known in astrophysics,  masses of the BSs' components represent one of the crucial parameters,  either for being a common parameter between the dynamical method and the spectrophotometrical one, or for being one of the most important results of the analysis for their essential role in stellar evolution theories.

The analysis of the the systems under study showed a good consistency between the mass sum using the dynamical solutions and the masses estimated using Al-Wardat's method. But, in order to achieve a better consistency,  a new dynamical parallax is given in this work for both systems.

%That is why we focused on this parameter, and hereafter we summarize the final results for each system:

%\textbf{HIP 19206}; The dynamical analysis give a total mass of  $M=1.9926\pm0.0620$\, $\rm{M_\odot}$, while the other

%\textbf{HIP 84425}; star has a higher mass of $M=2.8739\pm0.0492$\, $\rm{M_\odot}$.

The results show that the components of HIP 19206 system are main sequence stars, and the components of Hip 84425 have left the main sequence to the sub-giant stage. The components of each system have almost similar parameters and same age. Hence, fragmentation is the most probable process for the formation and evolution of both systems, where \cite{1994ASPC...65..115B} concluded that fragmentation of rotating disk around an incipient central protostar is possible, as long as there is continuing in fall, and \cite{2001IAUS..200.....Z} pointed out that hierarchical fragmentation during rotational collapse has been invoked to produce binaries and multiple systems.

 As consequence,  the  results of such studies would   improve our understanding of the nature, formation and  evolution of binary and multiple stellar systems.

Finally,  synthetic photometry, consistency of the results between the two methods and consistency between some parameters with the available observational data give an indication about the accuracy of the used methods, and hence the estimated stellar parameters for both systems.

%As a result, we  have strong grounds to determine the entire and individual synthetic magnitudes and color indices for both systems using different photometric systems such as Johnson, Str\"{o}mgren, Hipparcos and Tycho.

%As  illustrated in Figures 1 \& 2, the integrated motion with Tokovinin's dynamic method, helps us to solve the orbits of both systems and reflects in their final shapes due to the mass interaction in binaries.

%\section*{References}
%
%\bibliography{references.bib}
%
%\end{document}
%
%%\bibliographystyle{aa}
%\bibliographystyle{acta}
%%
%\bibliography{references}
%
%%
%\end{document}
%

\section{ACKNOWLEDGMENTS}

The data that support the findings of this study are openly available in The Gaia Data Release 2 (Gaia DR2 ) at https://gea.esac.esa.int/archive. This research has made use of SAO/NASA, SIMBAD database, Fourth Catalog of Interferometric Measurements of Binary Stars, Sixth Catalog of Orbits of Visual Binary Stars, IPAC data systems,  ORBITX code and Al-Wardat's complex method for analyzing close visual binary and multiple systems with its codes.

%\bibliographystyle{raa}
%\bibliography{References}

\begin{thebibliography}{41}
\providecommand\natexlab[1]{#1}
\providecommand\JournalTitle[1]{#1}

\bibitem[{Al-Wardat}(2002)]{2002BSAO...53...51A}
{Al-Wardat}, M.~A. 2002, Bulletin of the Special Astrophysics Observatory, 53,
  51

\bibitem[{Al-Wardat}(2007)]{2007AN....328...63A}
{Al-Wardat}, M.~A. 2007, Astronomische Nachrichten, 328, 63

\bibitem[{Al-Wardat}(2009)]{2009AN....330..385A}
{Al-Wardat}, M.~A. 2009, Astronomische Nachrichten, 330, 385

\bibitem[{Al-Wardat}(2012)]{2012PASA...29..523A}
{Al-Wardat}, M.~A. 2012, \pasa, 29, 523

\bibitem[{Al-Wardat}(2014)]{2014AstBu..69..454A}
{Al-Wardat}, M.~A. 2014, Astrophysical Bulletin, 69, 454

\bibitem[{Al-Wardat} {et~al.}(2014{\natexlab{a}})]{2014AstBu..69...58A}
{Al-Wardat}, M.~A., {Balega}, Y.~Y., {Leushin}, V.~V., {et~al.}
  2014{\natexlab{a}}, Astrophysical Bulletin, 69, 58

\bibitem[Al-Wardat {et~al.}(2016)]{al2016physical}
Al-Wardat, M.~A., El-Mahameed, M.~H., Yusuf, N.~A., Khasawneh, A.~M., \& Masda,
  S.~G. 2016, Research in Astronomy and Astrophysics, 16, 166

\bibitem[Al-Wardat {et~al.}(2021)]{al-wardat_hussein_al-naimiy_barstow_2021}
Al-Wardat, M.~A., Hussein, A.~M., Al-Naimiy, H.~M., \& Barstow, M.~A. 2021,
  Publications of the Astronomical Society of Australia, 38, e002

\bibitem[{Al-Wardat} {et~al.}(2014{\natexlab{b}})]{2014PASA...31....5A}
{Al-Wardat}, M.~A., {Widyan}, H.~S., \& {Al-thyabat}, A. 2014{\natexlab{b}},
  \pasa, 31, e005

\bibitem[Al-Wardat {et~al.}(2017)]{al2017physical}
Al-Wardat, M., Docobo, J., Abushattal, A., \& Campo, P. 2017, Astrophysical
  Bulletin, 72, 24

\bibitem[{Balega} {et~al.}(2006{\natexlab{a}})]{2006BSAO...59...20B}
{Balega}, I.~I., {Balega}, A.~F., {Maksimov}, E.~V., {et~al.}
  2006{\natexlab{a}}, Bull.~Special Astrophys.~Obs., 59, 20

\bibitem[{Balega} {et~al.}(2006{\natexlab{b}})]{2006AA...448..703B}
{Balega}, I.~I., {Balega}, Y.~Y., {Hofmann}, K.-H., {et~al.}
  2006{\natexlab{b}}, \aap, 448, 703

\bibitem[{Balega} {et~al.}(2007)]{2007AstBu..62..339B}
{Balega}, I.~I., {Balega}, Y.~Y., {Maksimov}, A.~F., {et~al.} 2007,
  Astrophysical Bulletin, 62, 339

\bibitem[{Bonnell}(1994)]{1994ASPC...65..115B}
{Bonnell}, I.~A. 1994, in Astronomical Society of the Pacific Conference
  Series, Vol.~65, Clouds, Cores, and Low Mass Stars, ed. {D.~P.~Clemens \&
  R.~Barvainis}, 115

\bibitem[Brown {et~al.}(2020)]{brown2020gaia}
Brown, A.~G., Vallenari, A., Prusti, T., {et~al.} 2020, arXiv preprint
  arXiv:2012.01533

\bibitem[Collaboration {et~al.}(2018)]{gaia2018vizier}
Collaboration, G., {et~al.} 2018, VizieR Online Data Catalog, 1345

\bibitem[{ESA}(1997)]{1997yCat.1239....0E}
{ESA}. 1997, {The Hipparcos and Tycho Catalogues (ESA)}

\bibitem[{Girardi} {et~al.}(2000)]{2000A&AS..141..371G}
{Girardi}, L., {Bressan}, A., {Bertelli}, G., \& {Chiosi}, C. 2000, \aaps, 141,
  371

\bibitem[Girardi {et~al.}(2000)]{girardi2000evolutionary}
Girardi, L., Bressan, A., Bertelli, G., \& Chiosi, C. 2000, Astronomy and
  Astrophysics Supplement Series, 141, 371

\bibitem[{Girardi} {et~al.}(2000)]{2000yCat..41410371G}
{Girardi}, L., {Bressan}, A., {Bertelli}, G., \& {Chiosi}, C. 2000, VizieR
  Online Data Catalog, 414, 10371

\bibitem[{Gray}(2005)]{2005oasp.book.....G}
{Gray}, D.~F. 2005, {The Observation and Analysis of Stellar Photospheres}, 505

\bibitem[{Hartkopf} {et~al.}(2001)]{2001AJ....122.3480H}
{Hartkopf}, W.~I., {McAlister}, H.~A., \& {Mason}, B.~D. 2001, \aj, 122, 3480

\bibitem[{Horch} {et~al.}(2008)]{2008AJ....136..312H}
{Horch}, E.~P., {van Altena}, W.~F., {Cyr}, William~M., J., {et~al.} 2008, \aj,
  136, 312

\bibitem[{Kurucz}(1994)]{1994KurCD..19.....K}
{Kurucz}, R. 1994, Solar abundance model atmospheres for 0,1,2,4,8 km/s.~Kurucz
  CD-ROM No.~19.~ Cambridge, Mass.: Smithsonian Astrophysical Observatory,
  1994., 19

\bibitem[{Lallement} {et~al.}(2018)]{2018AA...616A.132L}
{Lallement}, R., {Capitanio}, L., {Ruiz-Dern}, L., {et~al.} 2018, \aap, 616,
  A132

\bibitem[Lang(1992)]{lang1992astrophysical}
Lang, K. 1992, Astrophysical Data: Planets and Stars, 132ff

\bibitem[{Lang}(1992)]{1992adps.book.....L}
{Lang}, K.~R. 1992, {Astrophysical Data I. Planets and Stars.}, 133

\bibitem[{Ma{\'{\i}}z Apell{\'a}niz}(2007)]{2007ASPC..364..227M}
{Ma{\'{\i}}z Apell{\'a}niz}, J. 2007, in Astronomical Society of the Pacific
  Conference Series, Vol. 364, The Future of Photometric, Spectrophotometric
  and Polarimetric Standardization, ed. C.~{Sterken} (San Francisco:
  Astronomical Society of the Pacific), 227

\bibitem[Masda {et~al.}(2019)]{masda2019physical}
Masda, S., Docobo, J., Hussein, A., {et~al.} 2019, Astrophysical Bulletin, 74,
  464

\bibitem[Masda {et~al.}(2016)]{masda2016physical}
Masda, S.~G., Al-Wardat, M.~A., Neuh{\"a}user, R., \& Al-Naimiy, H.~M. 2016,
  Research in Astronomy and Astrophysics, 16, 112

\bibitem[{Masda} {et~al.}(2019)]{2019AstBu..74..464M}
{Masda}, S.~G., {Docobo}, J.~A., {Hussein}, A.~M., {et~al.} 2019, Astrophysical
  Bulletin, 74, 464

\bibitem[{Mason} {et~al.}(2018)]{2018yCat..51550215M}
{Mason}, B.~D., {Hartkopf}, W.~I., {Miles}, K.~N., {et~al.} 2018, VizieR Online
  Data Catalog, J/AJ/155/215

\bibitem[{Rica Romero}(2010)]{2010RMxAA..46..263R}
{Rica Romero}, F.~M. 2010, Revista Mexicana de Astronomía y Astrofísica, 46, 263

\bibitem[{Shatskii} \& {Tokovinin}(1998)]{1998AstL...24..673S}
{Shatskii}, N.~I., \& {Tokovinin}, A.~A. 1998, Astronomy Letters, 24, 673

\bibitem[S{\"o}derhjelm(1999)]{soderhjelm1999visual}
S{\"o}derhjelm, S. 1999, Astronomy and Astrophysics, 341, 121

\bibitem[{Tokovinin}(2019)]{2019IAUDS.199....1C}
{Tokovinin}, A. 2019, IAU Commission on Double Stars, 199, 2

\bibitem[{Tokovinin} {et~al.}(2010)]{2010AJ....139..743T}
{Tokovinin}, A., {Mason}, B.~D., \& {Hartkopf}, W.~I. 2010, \aj, 139, 743

\bibitem[{Tokovinin} {et~al.}(2015)]{2015AJ....150...50T}
{Tokovinin}, A., {Mason}, B.~D., {Hartkopf}, W.~I., {Mendez}, R.~A., \&
  {Horch}, E.~P. 2015, \aj, 150, 50

\bibitem[{Tokovinin} {et~al.}(2016)]{2016AJ....151..153T}
{Tokovinin}, A., {Mason}, B.~D., {Hartkopf}, W.~I., {Mendez}, R.~A., \&
  {Horch}, E.~P. 2016, \aj, 151, 153

\bibitem[{van Leeuwen}(2007)]{2007A&A...474..653V}
{van Leeuwen}, F. 2007, \aap, 474, 653

\bibitem[{Zinnecker} \& {Mathieu}(2001)]{2001IAUS..200.....Z}
{Zinnecker}, H., \& {Mathieu}, R., eds. 2001, IAU Symposium, Vol. 200, {The
  Formation of Binary Stars}

\end{thebibliography}

\end{document}